\documentclass[11pt]{article}
\usepackage{amsmath, amssymb, amscd, amsthm, amsfonts}
\usepackage{graphicx}
\usepackage{multirow}
\usepackage[table,xcdraw]{xcolor}
\usepackage{microtype}
\usepackage{abstract}
\usepackage{tocloft} 
\usepackage{tocbibind}
\usepackage{fancyhdr} 
\usepackage{titlesec}
\usepackage{float}
\usepackage{setspace}
\usepackage[plainpages=false,pdfpagelabels,unicode]{hyperref}
\usepackage{emptypage}
\usepackage{graphicx}
\usepackage{listings}
\usepackage{etoolbox}
\usepackage{xstring}
\usepackage[table]{xcolor}
\usepackage{pgffor}
\usepackage{nomencl}
\usepackage{esvect}
\usepackage{eso-pic}
\usepackage[export]{adjustbox}
\usepackage{tikz}
\usepackage{booktabs}
\usepackage{todo}
\usepackage{longtable}
\usepackage{verbatim} 	
\usepackage{pgfplots}
\usepackage{enumitem}
\usepackage{empheq}
\microtypecontext{spacing=nonfrench}
\graphicspath{{./Gambar/}}
\usepackage{caption}
\usepackage{subcaption}

\title{Bayesian Chain Ladder For Cumulative Run-Off Triangle Under Half-Normal Distribution Assumption}
\author{Rizky Reza Fauzi \and Jerremy Joelnathan Stevanlim}
\date{}

\pgfplotsset{compat=1.18}
\begin{document}

\maketitle

\begin{abstract}
An insurance company is required to prepare a certain amount of money, called reserve, as a mean to pay its policy holders’ claims in the future. There are several types of reserve, one of them is IBNR reserve, for which the payments are made not in the same calendar year as the events that trigger the claims. Wuthrich and Merz (2015) developed a Bayesian Chain Ladder method for gamma distribution that applies the Bayesian Theory into the usual Chain Ladder Method \cite{Merz:15:stochastic}. In this article, we modify the previous Bayesian Chain Ladder for half-normal distribution, as such distribution is more well suited for lighter tail claims.
\end{abstract}

\section{INTRODUCTION}
As a mean to pay future claims, an insurance company is required to prepare a certain amount of money called reserve. There several types of reserve, such as immediate reserve, for which the money is prepared for near future, reported but not settled or RBNS reserve, and incurred but not reported or IBNR reserve. In this article, we are focusing ourselves for IBNR reserve, where the reserved money is prepared for the claims that cannot be paid in the same calendar year as the events that trigger the claims.

The IBNR data usually is served in the form of a run-off triangle, a table for which its rows represent its accident year, and its columns are for its development years. From run-off triangle data, a method called Mack Chain Ladder (or just chain ladder for short) \cite{Thomas:93:MCL} is widely used to forecast the future IBNS claims. However, the chain ladder is a deterministic method, which means it does not consider the randomness of the claims.

To improve the performance of the chain ladder method, the Bayesian Chain Ladder method \cite{peters:17:bayesian} was introduced by fusing the Bayesian Statistical Theory into the chain ladder, specifically for a run-off triangle data with gamma distribution as the ruling distribution. Though gamma distribution is quite versatile as its tail weight ranges from light to heavy, but as light as it can be, the tail of gamma distribution cannot reach the same lightness of the tail of normal distribution. Hence, if the run-off triangle data has a much lighter tail, using Bayesian Chain Ladder with gamma distribution might lead to an overestimated reserve.

In this article, we are introducing a new type of Bayesian Chain Ladder for lighter tail run-off triangle data, specifically we are using the half-normal distribution. Since the half-normal distribution is a normal distribution with $\mu=0$ folded along the y-axis, it has the same tail lightness as the normal one. Thus, when dealing with a much lighter tail run-off triangle data, our method will not overestimate its IBNR reserve.

\section{LITERATURE REVIEW}
\subsection{Mack Chain Ladder Method}
Run-off triangle is a simplification a collective aggregate claim data from several individual claims data set \cite{antonio:06:logn}. The triangle is divided into two parts, which are the development triangle for where the paid claims recorded, and the future triangle, where the forecasting of the future claims should be made \cite{Olofsson:06:Business}. A run-off triangle can be in the form of incremental or cumulative. If the entry of the $i$-th row and $j$-th column, say $C_{i,j}$, represents the amount of claim that happened in the $i$-th accident year but paid after $j$ years, then the triangle is an incremental one. For the cumulative triangle, the entry, say $S_{i,j}$, represents the total amount of claims that happened in the $i$-th accident year, but paid $j$ years later at the latest. Obviously, here we have
\begin{equation}
    S_{i,j} = \sum_{k=0}^{j}C_{i,k},
\end{equation}
for $i=1,2,\dots,n$ and $j=0,1,\dots,n-1$.
\begin{table}[h]
\centering
\caption{Illustration of Cumulative Run-Off Triangle}
\label{tab:kumulatif}
\begin{tabular}{ccccccc}
\hline
 & \multicolumn{6}{c}{Tahun Penundaan} \\ \cline{2-7} 
\multirow{-2}{*}{Tahun Kejadian} & 0 & 1 & $\ldots$ & $j$ & $\ldots$ & $n$-1 \\ \hline
1 & $S_{1,0}$ & $S_{1,1}$ & $\ldots$ & $S_{1,j}$ & $\ldots$ & \cellcolor[HTML]{9698ED}$S_{1,n-1}$ \\
2 & $S_{2,0}$ & $S_{2,1}$ & $\ldots$ & $S_{2,j}$ & \cellcolor[HTML]{9698ED}$\ldots$ & \cellcolor[HTML]{FD6864}$\hat{S}_{2,n-1}$ \\
$\vdots$ & $\vdots$ & $\vdots$ & $\ddots$ & \cellcolor[HTML]{9698ED}$\vdots$ & \cellcolor[HTML]{FCFF2F}$\ddots$ & \cellcolor[HTML]{FD6864}$\vdots$ \\
$i$ & $S_{i,0}$ & $S_{i,1}$ & \cellcolor[HTML]{9698ED}$\ldots$ & \cellcolor[HTML]{FCFF2F}$\hat{S}_{i,j}$ & \cellcolor[HTML]{FCFF2F}$\ldots$ & \cellcolor[HTML]{FD6864}$\hat{S}_{i,n-1}$ \\
$\vdots$ & $\vdots$ & \cellcolor[HTML]{9698ED}$\vdots$ & \cellcolor[HTML]{FCFF2F}$\ddots$ & \cellcolor[HTML]{FCFF2F}$\vdots$ & \cellcolor[HTML]{FCFF2F}$\ddots$ & \cellcolor[HTML]{FD6864}$\vdots$ \\
$n$ & \cellcolor[HTML]{9698ED}$S_{n,0}$ & \cellcolor[HTML]{FCFF2F}$\hat{S}_{n,1}$ & \cellcolor[HTML]{FCFF2F}$\ldots$ & \cellcolor[HTML]{FCFF2F}$\hat{S}_{n,j}$ & \cellcolor[HTML]{FCFF2F}$\ldots$ & \cellcolor[HTML]{FD6864}$\hat{S}_{n,n-1}$ \\ \hline
\end{tabular}
\end{table}
The entries of the development triangle (upper-left side) are indexed by $i=1,2,\dots,n$ and $j=0,1,\dots,n-i$ , while the indexes of the entries in the future triangle are $i=2,3,\dots,n$ and $j=n-i+1,n-i+2,\dots,n$.

In the chain ladder method, forecasting is made by calculating the development factors of each development year from the cumulative triangle \cite{Brown:08:MCL}. There are some assumptions for this method, which are
\begin{itemize}
    \item The random variables $S_{i,j}$’s are independent through different accident years, 
    \item For each development year, there is a development factor $f_j>0$ for which we assume $S_{i,j}=f_jS_{i,j-1}$. 
\end{itemize}

Mack (1993) estimated the development factors with
\begin{equation}
    \hat{f}_j^{CL} = \frac{\sum_{i=1}^{n-j}S_{i,j}}{\sum_{i=1}^{n-j}S_{i,j-1}},
\end{equation}
for $j=1,2,\dots,n-1$. Hence, the ultimate claim, i.e. the cumulative claim in the last development year, can be estimated by 
\begin{equation}
    \hat{S}_{i,n-1}^{CL} = S_{i,n-i}\prod_{j=n-i+1}^{n-1}\hat{f}_j^{CL},
\end{equation}
for $i=2,3,\dots,n$. Next, the outstanding claim for each accident year can be calculated with 
\begin{equation}
    \hat{R}_i^{CL} = \hat{S}_{i,n-1} - S_{i,n-i},
\end{equation}
for $i=2,3,\dots,n$, and the IBNR reserve would be
\begin{equation}
    \hat{R}^{CL} = \sum_{i=2}^{n} \hat{R}_i^{CL}.
\end{equation}
\subsection{Half-Normal Distribution}
Let $Y \sim N\left(0,\frac{\omega\pi}{2}\right)$ and define $X=|Y|$, then $X$ has a half-normal distribution with scale parameter , denoted by $X\sim half.N(\omega)$. By change of variable technique \cite{hog:19:stat}, the probability density function of the half-normal distribution is
\begin{equation}
    f_X(x) = \frac{2}{\pi\sqrt{\omega}}e^\frac{-x^2}{\omega\pi},
\end{equation}
for $0<x<\infty$.
\begin{figure}[hbt!]
    \centering
    \includegraphics[page=3,width=.7\textwidth]{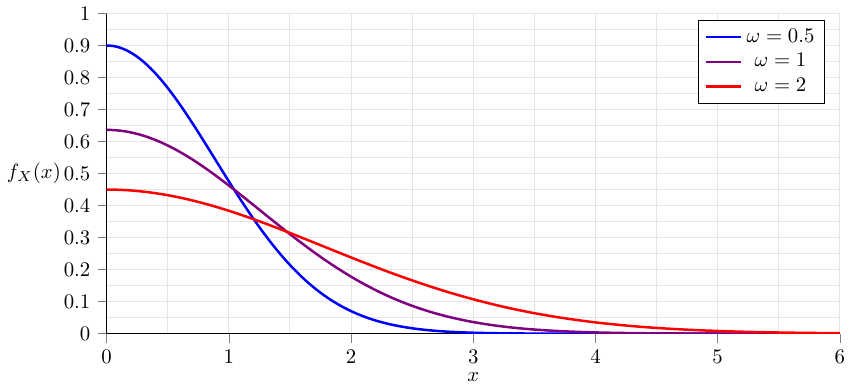}
    \caption{Graph of Half-Normal Density }
    \label{fig:pdf halfnormal}
\end{figure}

From \ref{fig:pdf halfnormal}, we can see that the shape of half-normal density graph is really just the right half side of the normally bell curve with $\mu=0$. Also, the role of the scale parameter is similar to the variance of the normal distribution, which means the larger the value of the scale, the flatter the graph is.

Using substitution technique, it can be easily proven that
\begin{equation}
    E(X)=\sqrt{\omega},
\end{equation}
and by the property of $E(Y^2)=\frac{\pi\omega}{2}$, we have
\begin{equation}
    E(X^2)=\frac{\pi\omega}{2},
\end{equation}
that leads to
\begin{equation}
    \mathrm{Var} (X)=\omega\left(\frac{\pi}{2}-1\right),
\end{equation}
which resonates to the interpretation of the half-normal curve.

The moment generating function of half-normal distribution can be determined by utilizing a similar maneuver to find the normal moment generating function, and we get
\begin{equation}
    M_X(t)=2\exp\left\{\frac{\pi\omega}{4}t^2\right\}\left[1-\Phi\left(-t\sqrt{\frac{\pi\omega}{2}}\right)\right],
\end{equation}
for $-\infty<t<\infty$. From here, we can conclude that all of the positive raw moments of the half-normal distribution exist, meaning that the tail weight is light enough \cite{Klugman:08:Klugman}.

\section{MATERIAL AND METHOD} \label{material}
\subsection{Bayesian Chain Ladder for Half-Normal Run-Off Triangle}
 The essence of Bayesian Chain Ladder is applying Bayesian Statistical Theory into chain ladder method to forecast the IBNR reserve \cite{Bayesian:17:William}. The concept of Bayesian Theory is used in the construction of the development factors, which we will call Bayesian development factors \cite{Lambert:18:student}. Here we assume that the cumulative run-off triangle is ruled under the half-normal distribution, as we are trying to introduce a new Bayesian method for light tail run-off triangle data. Next, we need to construct a suitable prior distribution, which we choose to utilize the inverse gamma distribution. At last, after determining the posterior distribution, we can find the formula of the Bayesian development factor.

Hence, the assumptions we need to construct new Bayesian development factors are as follows
\begin{itemize}
    \item From the development year point of view, the cumulative claims $S_{i,j}$ follow the Markov Property, which is
    \begin{equation}
        (S_{i,j-1},S_{i,j-2},\dots,S_{i,0})=(S_{i,j-1}),
    \end{equation}
    \item The prior random variable $\mathbf{\Theta}=(\Theta_1,\Theta_2,\dots,\Theta_{n-1})$ are independent and each one of them follows inverse gamma distribution (Wuthrich \& Merz, 2008), or mathematically, 
\begin{equation}
     \Theta_j \sim \text{inv.}\Gamma(\alpha_j,\beta_j),
\end{equation}
    \item The cumulative claim $S_{i,j}$ is independent to the prior variable $k$ if $k\neq j$, i.e. 
\begin{equation}
    (\boldsymbol{\theta})=(\theta_j)
\end{equation}
\item The cumulative claims in the $0$-th development year $S_0=S_{1,0},S_{2,0},\dots,S_{n,0}$ are independent with joint density $f_{S_0}$.
\item  We assume
\begin{equation*}
    (S_{i,j}|S_{i,j-1},\theta_j)\sim \textit{half.}N(\theta_j S_{i,j-1}^2)
\end{equation*}
\end{itemize}
 
From those assumptions, we will have 
\begin{equation}
    E(S_{i,j-1},S_{i,j-2},\dots,S_{i,0},\theta)=E(S_{i,j-1},\theta_j)=S_{i,j-1}\sqrt{\theta_j},
\end{equation}
which means that the Bayesian development factor is based on $\sqrt{\theta_j}$. Furthermore, since 
\begin{equation}\label{eq: ekspetasi akar}
    E(\sqrt{\theta_j})=\frac{\Gamma(\alpha_j-\frac{1}{2})}{\Gamma(\alpha_j)} \sqrt{\beta_j},
\end{equation}
the prior development factor is based on equation \eqref{eq: ekspetasi akar}. 

By the independence \cite{hoff:09:bayesian}, we have the prior density is
\begin{equation}
    \pi_\Theta(\theta) \propto \prod_{j=1}^{n-1} \frac{1}{\theta_j^{\alpha_j+1}}e^{-\frac{\beta_j}{\theta_j}}
\end{equation}
while the model is
\begin{equation}
    f_{S|\Theta}(\theta) \propto \prod_{j=1}^{n-1} \frac{1}{\theta_j^{\frac{1}{2}(n-j)}}\exp\left\{-\frac{1}{\pi\theta_j}\sum_{i=1}^{n-j}\frac{S_{i,j}^2}{S_{i,j-1}^2}\right\}.
\end{equation}
Therefore, we have their joint posterior density is
\begin{equation}
    \pi_{\Theta|S}(\theta|s)\propto \prod_{j=1}^{n-1} \frac{1}{\theta_j^{\alpha_j+\frac{1}{2}(n-j)+1}}\exp\left\{-\frac{1}{\theta_j}\left(\beta_j+\frac{1}{\pi}\sum_{i=1}^{n-j}\frac{S_{i,j}^2}{S_{i,j-1}^2}\right)\right\},
\end{equation}
which leads to the conclusion that
\begin{equation}
    \Theta_j|S\sim inv.\Gamma\left(\alpha_j+\frac{1}{2}(n-j),\beta_j+\frac{1}{\pi}\sum_{i=1}^{n-j}\frac{S_{i,j}^2}{S_{i,j-1}^2}\right).
\end{equation}
Thus, the formula of our Bayesian development factor is
\begin{equation}
    \hat{f}_j^B = E(\sqrt{\Theta_j}|S)= \frac{\Gamma\left(\alpha_j + \frac{1}{2}(n-j-1)\right)}{\Gamma\left(\alpha_j + \frac{1}{2}(n-j)\right)} \sqrt{\beta_j+\frac{1}{\pi}\sum_{i=1}^{n-j}\frac{S_{i,j}^2}{S_{i,j-1}^2}},
\end{equation}
for $j=1,2,\dots,n-1$.

Furthermore, the Bayesian ultimate claim can be calculated by
\begin{equation}
    \hat{S}_{i,n-1}^B = S_{i,n-i} \prod_{j=n-i+1}^{n-1} \hat{f}_j^B,
\end{equation}
for $i=2,3,\dots,n$. Next, the Bayesian outstanding claim for each accident year can be calculated with 
\begin{equation}
    \hat{R}_i^B = \hat{S}_{i,n-1}^B - \hat{S}_{i,n-i},
\end{equation}
for $i=2,3,\dots,n$, and the Bayesian IBNR reserve would be 
\begin{equation}
    \hat{R}^B = \sum_{i=2}^{n} \hat{R}_i^B.
\end{equation}
\subsection{Compulsory Third Party Insurance Data}
The data we are using is a secondary data from the annual report of Australian Prudential Regulation Authority (APRA) with title “General Insurance Claims Development Statistics December 2021”. Here we are analyzing the gross IBNR claims from the compulsory third party (CTP) insurance data that have been paid from 2012 until 2021 in Australian dollar. To compare the performance of our new method, first we will show the IBNR forecasting results using the chain ladder method. Then, with the same data, we will calculate the results if we use the half-normal Bayesian Chain Ladder.

 \section{RESULTS AND DISCUSSION}
 The results from the Mack Chain Ladder method are as follows.
 \begin{figure}
     \centering
     \includegraphics[width=0.85\linewidth]{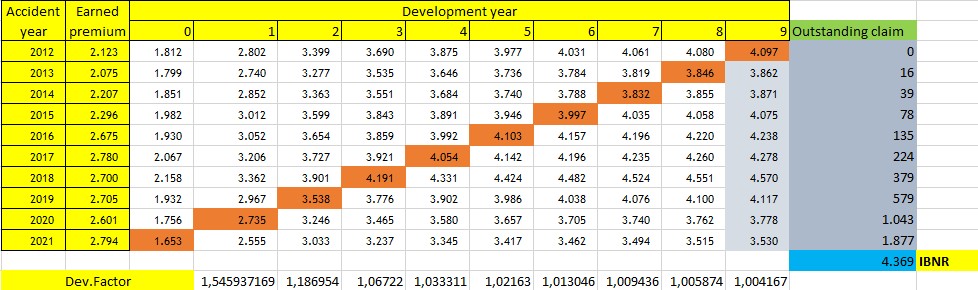}
     \caption{Mack Chain Ladder Results}
     \label{fig:mcl}
 \end{figure}
 As we can see from Figure \ref{fig:mcl}, the IBNR reserve prediction based on chain ladder method is 4,359 Australian dollars, with decreasing development factors through the development years. This is quite logical the increments of the cumulative claims are decreasing as well. However, the decreasing increment is one of the indicators that the distribution of the data has a lighter tail. Hence, this IBNR reserve might be overestimated.

 In case of our new method, we chose to set $\alpha_j=45$ for each development year. The reason is related to the interpretation of the shape parameter $\alpha_j$ itself. It is clear that the inverse gamma dan inverse chi-square distributions are related (so are gamma and chi-square), where the shape parameter of inverse gamma distribution is related to the degrees of freedom. Since we have to predict 45 entries, we can say the degrees of freedom of this data is 45. Furthermore, related to equation \eqref{eq: ekspetasi akar} and the chain ladder development factor, we choose
 \begin{equation}
     \beta_j = \frac{\Gamma^2(\alpha_j)\sum_{i=1}^{n-j}S_{i,j}^2}{\Gamma^2(\alpha_j-\frac{1}{2})\sum_{i=1}^{n-j}S_{i,j-1}^2}.
 \end{equation}
 Thus, the IBNR reserve forecasted by the new half-normal Bayesian Chain Ladder is smaller, as showed as follows.
 \begin{figure}
     \centering
     \includegraphics[width=0.85\linewidth]{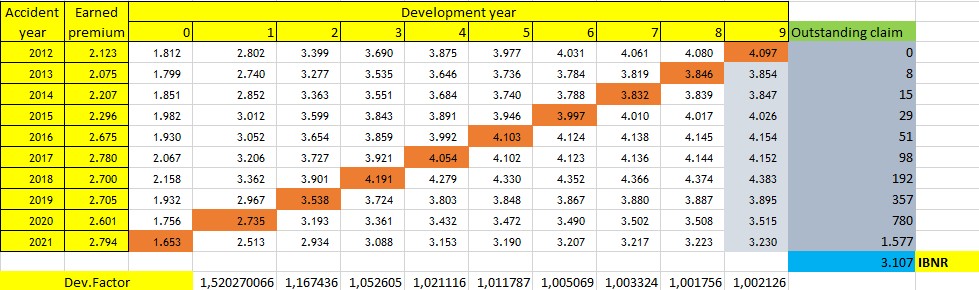}
     \caption{Half-Normal Bayesian Chain Ladder Results }
     \label{fig:halfn}
 \end{figure}
It is clear that there is a glaring difference between the results, as the IBNR reserve from our method is significantly smaller. Though the behavior of the Bayesian development factors is similar (decreasing), we witness a more drastic shrinking here. This actually resonates with the half-normal distribution, for which its tail decreases more rapidly than the others, implying the development factors should get smaller quicklier.

\section{CONCLUSION}
Based on the analysis we have done, half-normal Bayesian Chain Ladder is a method that applies Bayesian Theory to the usual chain ladder method, under the assumption that the cumulative claims have a half-normal distribution, a really light tail distribution. Bayesian development factor formula we have derived here is based on the combination of the model information, which is half-normal, with the prior knowledge inverse gamma, results in a posterior information inverse gamma. After the data analysis, we found that the IBNR reserve calculated by our new method is much smaller than the chain ladder one, as our new method captures the light tail property of the data.\\
\\
\noindent\large{\textbf{ACKNOWLEDGEMENT}}\\
The authors are really grateful for the kind reviews and inputs from the reviewers, that enhance this article greatly.

\bibliographystyle{compj.sty} 
\bibliography{references.bib}

\begin{thebibliography}{99}

\bibitem{Merz:15:stochastic}
Wüthrich, M.~V. dan Merz, M. (2015) {Stochastic claims reserving manual: advances in dynamic modeling}.
\newblock {\em SSRN Electronic Journal}, {\bf  15-34}, 11--54.

\bibitem{Thomas:93:MCL}
Mack, T. (1993) {Distribution-free calculation of the standard error of chain ladder reserve estimates}.
\newblock {\em ASTIN Bulletin}, {\bf  23(2)}, 213–225.

\bibitem{peters:17:bayesian}
Peters, G.~W., Targino, R.~S., dan Wüthrich, M.~V. (2017) {Full Bayesian analysis of claims reserving uncertainty}.
\newblock {\em Insurance: Mathematics and Economics}, {\bf  73}, 41--53.

\bibitem{antonio:06:logn}
Antonio, K., Beirlant, J., Hoedemakers, T., dan Verlaak, R. (2006) {Lognormal Mixed Models For Reported Claims Reserves}.
\newblock {\em North American Actuarial Journal}, {\bf  10(1)}, 30--48.

\bibitem{Olofsson:06:Business}
Olofsson, M. (2006) {Stochastic Loss Reserving Testing the New Guidelines from the Australian Prudential Regulation Authority (ARPA) on Swedish Portfolio Data Using a Bootstrap Simulation and the Distribution-free Method by Thomas Mack}.
\newblock {\em Examensarbete}, {\bf  13}.

\bibitem{Brown:08:MCL}
Brown, R.~L. dan Lennox, W.~S. (2015) {\em {Introduction to Ratemaking and Loss Reserving for Property and Casualty Insurance}},  4th edition. ACTEX Publications,  New Hartford.

\bibitem{hog:19:stat}
Hogg, R.~V., McKean, J.~W., dan Craig, A.~T. (2019) {\em {Introduction to Mathematical Statistics}},  8th edition. Pearson,  Boston.

\bibitem{Klugman:08:Klugman}
Klugman, S., Panjer, H., dan Willmot, G. (2019) {\em {Loss Models from Data to Decisions}},  5th edition. John Wiley and Sons, Inc.,  USA.

\bibitem{Bayesian:17:William}
Bolstad, W.~M. dan Curran, J.~M. (2017) {\em {Introduction to Bayesian Statistics}},  3rd edition. John Wiley and Sons, Inc.,  USA.

\bibitem{Lambert:18:student}
Lambert, B. (2018) {\em {A Student’s Guide to Bayesian Statistics}},  1st edition. SAGE Publications Ltd.,  California.

\bibitem{hoff:09:bayesian}
Hoff, P.~D. (2009) {\em {A First Course in Bayesian Statistical Methods}}. Springer Publishing,  New York.

\end{thebibliography}
\appendix

\end{document}